\documentclass{emulateapj}
\usepackage{apjfonts}
\usepackage{amsbsy}

\newcommand{\msun}{M_{\odot}}
\newcommand{\lsun}{L_{\odot}}
\newcommand{\kms}{{\rm km\, s}^{-1}}

\newcommand{\mpc}{\ {\rm Mpc}}
\newcommand{\hmpc}{{h^{-1}\mpc}}
\newcommand{\vv}{\upsilon}
\newcommand{\up}[1]{{\rm #1}}
\newcommand{\etal}{et al.\ }

\begin{document}

\title{Formation of the Black Holes in the Highest Redshift Quasars}

\author{Jaiyul Yoo and Jordi Miralda-Escud\'e}

\affil{Department of Astronomy, The Ohio State University, 
140 West 18th Avenue, Columbus, OH 43210; \\
jaiyul, jordi@astronomy.ohio-state.edu}

\slugcomment{accepted for publication in The Astrophysical Journal Letters}

\begin{abstract}
  The recent discovery of luminous quasars up to a redshift $z=6.43$ has
renewed interest in the formation of black holes massive enough to power
quasars. If black holes grow by Eddington-limited gas accretion with a
radiative efficiency of at least 10\%, the time required to grow from a
stellar black hole to $\sim 10^9 \msun$ is $\sim 10^9$ years, close to
the age of the universe at $z=6$. Black hole mergers may accelerate the
rate of mass growth, but can also completely eject black holes from halo
centers owing to the gravitational wave recoil effect. Recently, Haiman
concluded that black hole ejections likely do not allow black holes to
grow to $\sim 10^9 \msun$ by $z=6.43$. We reexamine this problem and
show that, by using a different halo escape velocity, 
accounting for the dependence of the recoil velocity on the
black hole binary mass ratio and spins, and allowing seed black holes to
form in all halos down to virial temperatures of 2000~K,
black hole masses may reach
$\sim 10^9 \msun$ as early as $z=9$ starting from stellar seeds, without
super-Eddington accretion. In this particular case, we find that these massive
black holes form from the merger of $\sim 10^4$ stellar black holes 
formed in low-mass halos at $z\sim 20$, 
which must all grow close to the maximum Eddington rate
over most of the time available from their birth to $z\sim 6$. The
alternative is that black holes can grow more rapidly by super-Eddington
accretion.
\end{abstract}


\section{Introduction}
\label{sec:int}

  Highly luminous quasars at high redshift require the formation of
black holes (BH) with a mass $M$ large enough to make the observed
luminosity be less than the Eddington limit. This may pose a problem for
cold dark matter (CDM) models where massive halos that can harbor BHs do not
form until late epochs \citep{efs}.
Currently, the highest redshift quasar known
is SDSS 1148+3251, at $z=6.43$ \citep{fan}. Its luminosity $L$, if
equated to the Eddington value $L_\up{Edd}=4\times10^4\lsun(M/\msun)$,
results in a minimum BH mass $M=4\times 10^9\msun$ \citep{fan,haiman}.
The quasar luminosity inferred from the observed flux is not likely to
be greatly overestimated due to gravitational lensing or beaming
\citep{hc02,willott,keeton}.

  If BHs grew in mass by accretion through a standard thin disk with
radiative efficiency $\epsilon \sim 0.1$, then a BH
of initial mass $M_0$ formed at time $t_0$ accreting continuously
up to time $t$ with luminosity $L$ will reach a final mass 
$M = M_0 \, \exp[(t-t_0)/t_\up{Sal}]$,
where $t_\up{Sal}$ is the Salpeter time \citep{salp},
\begin{equation}
t_\up{Sal}={\epsilon Mc^2\over (1-\epsilon)L}= 4\times 10^7 \,{\rm yr} \,
{\epsilon \over 0.1 (1-\epsilon)} \, {L_\up{Edd} \over L} ~.
\end{equation}
Note that the factor $1-\epsilon$ in the denominator is to account for
the fact that only a fraction $1-\epsilon$ of the accreted mass is
added to the BH.
Since the age of the universe at $z=6$ is $\sim 10^9$ years, 
a BH made by one of the 
first stars formed in the universe with $M_0\sim 10 \msun$ would need
to be accreting near the Eddington rate for almost all the time
available before $z=6$ in order to increase its mass by a factor
$\sim 10^8$ to power the observed quasars
(corresponding to $\sim 20$ e-folding times).

  There are several ways to ease this requirement and allow BHs to
grow to a large mass in less time: (a) BHs may have started at a
higher initial mass from the collapse of supermassive stars (e.g., 
\citealt{carr}).
 (b) There may be super-Eddington accretion with $t_\up{Sal} < 4\times 10^7$ 
years if the radiative efficiency is $\epsilon< 0.1$ 
(e.g., \citealt{ohsuga}), or
the luminosity is greater than the Eddington luminosity with $\epsilon=0.1$
\citep{rusz}. (c) Finally, several BHs may have merged to form a more
massive BH.

  In this paper we consider the latter possibility in the context of the
CDM model, where dark matter halos merge
hierarchically. If the earliest BHs arose from the first stars forming
in the center of halos, then halo mergers can lead to the merger of
their central BHs (e.g., \citealt{volon}), after dynamical friction
leads to the formation of a binary BH which can then lose orbital energy
by interacting with stars \citep{rees} or via gas dissipation
\citep{andy}, until the emission of gravitational waves takes over. If
many BHs growing from initial stellar seeds of mass $M_{0_i}$ made at
time $t_{0_i}$ continuously accrete until time $t$ (as they merge with
each other at any intermediate times), the final BH can reach a mass
\citep{hl},
\begin{equation}
 M = \sum_i M_{0_i} \, e^{(t-t_{0_i})/t_\up{Sal}} ~.
\label{eq:bhms}
\end{equation}

  In the last stage of the merger of BHs, the emission of gravitational
waves (GW) can give a net momentum to the resulting BH large enough to
eject it from its halo \citep{fitchett,kidder,favata,madau2,merritt}.
Then the halo center is left empty and the process of BH growth must
start over again. Recently, \citet{haiman} presented a calculation of
the growth of BH masses assuming that they start growing from stellar
seeds only when the halo has a velocity dispersion $\sigma > \vv_{kick}/
2$, where $\vv_{kick}$ is a fixed kick velocity given to the BHs by the
final GW burst. \citet{haiman} concluded that for $\vv_{kick} > 64~
\kms$, BHs would not be able to grow to large enough masses without
super-Eddington accretion.

\begin{figure}[t]
\centerline{\epsfxsize=3.5truein\epsffile{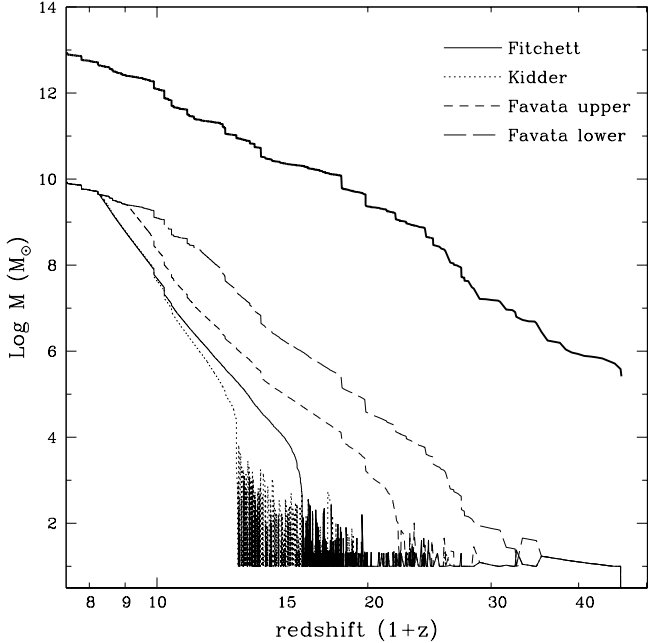}}
\caption{Mass of the main BH progenitor as a function of redshift ending
at $z=6.43$, according to four different estimates of GW recoil effect
(shown as the four lower curves). The
upper curve (thick solid line) shows the mass of the main halo
progenitor. Seed BHs are started at $10 \msun$, and the maximum BH mass
allowed is $10^{-3}$ of the halo mass. For reference, GW recoil
velocities are of order $1000$, $1000$, $400$, and $50~\kms$
for Fitchett, Kidder, Favata upper and lower models, respectively.}
\label{fig:mass}
\end{figure}

  We show in this paper that when the adequate kick velocity as a
function of the mass ratio and spins of the two merging BHs is used,
and a halo escape velocity based on an isothermal profile (higher than
assumed by Haiman) is used, ejections by GW emission result in a smaller
reduction than found by \citet{haiman} of the final black hole mass that
is attained. As a
result, we shall show that BH masses of $10^9 \msun$ can be reached up
to $z\simeq 9$ without super-Eddington accretion, even for the highest
plausible kick velocities found by \citet{favata}. Our model for the BH
evolution and growth is described in \S~2. The main results are
presented in \S~3, and we discuss the implications in \S~4.

\section{Black Hole Evolution Model}
\label{sec:evolution}

  We model the evolution of dark matter halos using the merger tree
of \citet{cole} based on the extended 
Press-Schechter formalism \citep{press,bcek,lacey}. We use the flat
$\Lambda$CDM power spectrum \citep{daniel} with matter density $\Omega_m=0.3$, 
baryon density $\Omega_b=0.04$, present Hubble constant $H_0=70~\kms\mpc^{-1}$,
power spectrum normalization $\sigma_8=0.9$, and primordial index $n=1$,
consistent with the WMAP results \citep{bennett,spergel}. 
We start the merger tree at $z=6.43$ with a final halo mass
$M_f$, finding all the progenitor halos at higher redshift. The mass
$M_f$ is determined by requiring the quasar number density $n_Q=2.7
\times10^{-9}(\hmpc)^3$ \citep{fan}
to be equal to the number density of halos of mass $M_f$
at $z=6.43$. This gives the maximum possible mass of the host halo of
the quasar consistent with the observed quasar abundance (for example,
if quasars have a short duty cycle the host halo abundance must be
higher than the quasar abundance, so the halo mass must be lower). The
resulting halo mass, $M_f \simeq 8.5\times 10^{12}\msun$ \citep{haiman},
requires a $5.2 \sigma$ fluctuation at $z=6.43$.

  The first BHs are assumed to form in halos that have reached a virial
temperature $T=2000$~K, when molecular hydrogen cooling induces the
formation of the first stars (e.g., \citealt{yosh}). A constant initial
BH mass $M_0$ is assumed in every halo reaching this virial temperature.
The mass corresponding to this virial temperature, $M_{res}(z)$, is
chosen as the resolution of the merger tree. In order to accurately
follow the appearance and growth of halos down to mass $M_{res}$, we
compute accretion into halos of mass $M_h$ from smaller halos down to a
minimum mass $10^{-3} M_h$ or $M_{res}$, whichever is smallest. We
compute the virial temperature of halos, related to velocity dispersion
as $T=\mu \sigma^2/k$ (where $k$ is Boltzmann's constant and $\mu$ is
the mean particle mass), 
using the fitting formula of \citet{bryan} for the mean
overdensity of virialized halos $\Delta_c$, and the equations $\sigma^2
=GM_h/(2R_h)$, $M_h =(4\pi/3) R_h^3 \rho_c(z) \Delta_c(z)$, where $R_h$
and $\rho_c$ are the radius of virialization of the halo and the
critical density at a given redshift $z$, respectively.

\begin{figure}[b]
\centerline{\epsfxsize=3.5truein\epsffile{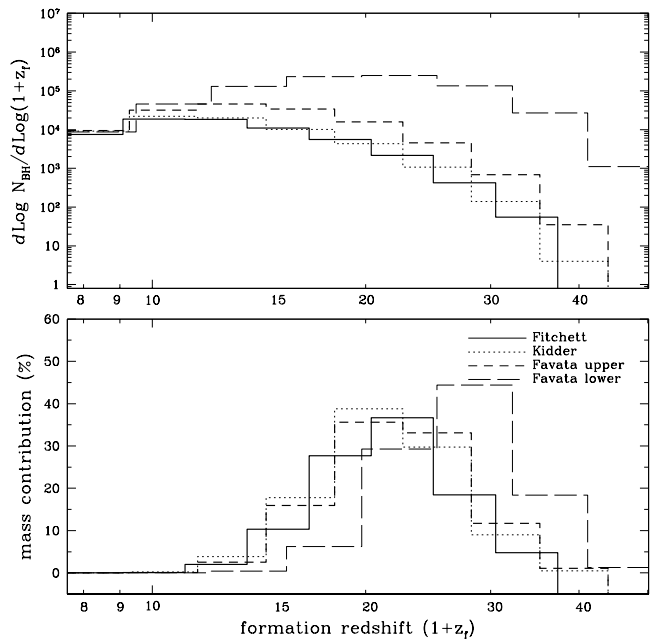}}
\caption{Composition of the final BH. {\it Upper panel:} Number of seed
BHs that merged into the final BH, as a function of their formation
redshift. {\it Lower panel:} Contribution to the final mass of the
BH from the BH seeds as a function of their formation redshift.}
\label{fig:hist}
\end{figure}

  We assume that all the BHs grow continuously by gas accretion on the
timescale $t_\up{Sal}$ obtained with $\epsilon/(1-\epsilon)= 0.1$ and
$L=L_\up{Edd}$, after their host halo has reached a minimum velocity
dispersion $\sigma_{min}$. We shall show some results for
$\sigma_{min}=0$, in which case BHs start growing as soon as they are
formed. Alternatively, the effects of slow gas cooling, stellar winds,
or supernovae may prevent any gas concentrating in the halo center and
being captured by the BH for accretion when $\sigma < \sigma_{min}$. The
BH growth by gas accretion continues until the BH mass reaches a
fraction $10^{-3}$ of its halo mass, at which point we simply set
$M = 10^{-3} M_h$.

  The last step is to check if the BH merger results in an ejection
from the host halo. We consider four estimates of the GW recoil
velocity. The first is a quasi-Newtonian calculation of non-spinning BHs
by \citet{fitchett}.
Second, \citet{kidder} added a post-Newtonian spin-orbit correction
to Fitchett's work that depends on the BH spins.
Lastly, \citet{favata} and \citet{merritt}
obtained a new estimate using BH perturbation theory, and
they provided a lower and upper limit on the GW kick velocity comprising
the uncertainty due to the final plunging state which gives the dominant
contribution (see their eqs.~[1],~[2]). We use these limits as two
separate estimates.

  The mass acccretion histories of BHs are calculated following the
merger trees in time and computing the GW recoil velocity every time two
halos merge (and by assumption, their central BHs merge). The
dimensionless spin parameters of merging BHs are uniformly chosen from 
antiparallel ($\hat a=-1$) to parallel ($\hat a=1$) in relation to the
orbital angular momentum, to calculate the effective spin parameters
\citep{damour1,damour2}. We then compare the GW kick velocity with the
escape velocity of the halo to decide if the merged BH is
ejected. We compute the escape velocity assuming the halo has an
isothermal profile, with density $\rho \propto r^{-2}$ down to the
radius of the BH zone of influence: $R_z\equiv GM/2\sigma^2$.
Therefore, the escape speed at $R_z$ is $\vv_{esc} = 2\sigma
\sqrt{1+\ln(R_h/R_z)}$, or about $5~\sigma$ when $M=10^{-3}M_h$.
Note that Haiman (2004) assumed instead that $\vv_{esc} = 2\sigma$, and
that no BHs start growing until the halo escape velocity has
exceeded an assumed, fixed GW kick velocity. The exact value of the
escape velocity depends on the halo density profile, but it would not be
much lower than we assume for a profile similar to observed galaxies.
  Gravitational waves emission also implies that a fraction of the mass
in BHs is lost as the energy of the waves. This loss of energy
does not greatly reduce the final BH mass \citep{menou},
and we neglect it in this paper.

\section{Results}
\label{sec:results}

  Figure~\ref{fig:mass} shows the mass history of the BH in the $z=6.43$
quasar SDSS~1148+5251 in our models. The upper thick solid line shows
the mass of the main halo progenitor in one realization of the merger
tree, obtained by always choosing the branch of the most massive
progenitor at every merger. The lower four lines show the mass of the BH
in this main halo progenitor, according to our four different
prescriptions for the GW kick velocity. The mass of the seed BHs is
fixed at $M_0 = 10 \msun$, and they start growing by gas accretion
immediately as they are formed, over a time $t_\up{Sal}=4\times 10^7$
years. At every halo merger, the mass of the two BHs is added when the
GW kick velocity is smaller than the escape velocity. This results in
the sudden BH mass increases seen in the figure. Growth by gas accretion
is assumed to continue immediately after mergers. When the GW kick
velocity exceeds the escape velocity, the BH is removed and replaced
immediately with a new seed with $M_0=10 \msun$. The figure shows that
BHs are often kicked out at high redshift when they reside in low-mass
halos, but this stops when the halo reaches a high enough escape
velocity. Large, sudden mass increases (for example, at $z\simeq 13$ for
the Kidder curve) are due to additions of BHs that have grown to high
mass in a merging subhalo. Note that the Favata lower limit curve
reaches a mass $10^{-3} M_h$ (at which point the BH is not allowed to
grow further) at $z\simeq 10$.

Figure~\ref{fig:hist} shows a histogram of 
the number of BHs that have merged into
the final BH at $z= 6.43$, as a function of their formation redshift
({\it top panel}), and their contribution to the total mass of the final
BH ({\it bottom panel}). In other words, the number shown in the
top panel is the number of BHs that we need to add over in equation
(\ref{eq:bhms}) from every interval of BH formation redshift, and the
bottom panel shows their contribution to the final mass including
the factor $e^{(t-t_{0_i})/t_\up{Sal}}$. The figure shows that the final mass
attained by the BH at $z=6.43$ is the result of the merger of many small
BHs: the formation redshift bins with the highest contribution to the
final BH mass typically contain $10^4$ to $10^5$ seed black holes that
merged. Hence, the $\sim 10^8$ mass growth factor from the initial
seed mass of $10 \msun$ to the final BH mass of $\sim 10^9 \msun$
originates in about equal parts from the growth factor
$e^{(t-t_{0_i})/t_\up{Sal}}$ of each individual BH, and from the number
of similar terms that are added together in equation (\ref{eq:bhms})
from all the BHs that have merged.

  In summary, these results show that in the optimistic model we have
assumed for BH growth,
it is possible to reach a mass as high as that inferred for the BH in
the quasar SDSS 1148+5251 at a redshift of up to $z\simeq 10$ for the
\citet{favata} lower limit of the GW kick velocity. Even for the highest
GW kick velocities the same BH mass can still be reached at $z\simeq 8$.

\begin{figure}[t]
\centerline{\epsfxsize=3.5truein\epsffile{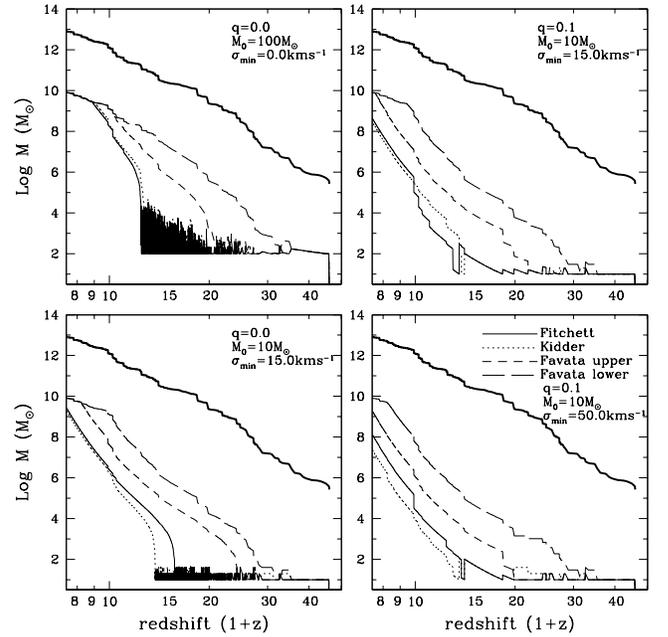}}
\caption{BH mass histories as in Fig. 1, changing model parameters for
mergers and growth by gas accretion. {\it Upper left panel:} the seed
BH mass is increased from $10 \msun$ to $100 \msun$. 
{\it Lower left panel:} Growth by gas accretion starts only once the
host halo has reached a minimum velocity dispersion 
$\sigma_{min}= 15~\kms$.
{\it Upper right panel:} BH mergers occur only when merging halos have
a mass ratio higher than $q=0.1$.
{\it Lower right panel:} The minimum velocity dispersion is increased
to $\sigma_{min} = 50~\kms$.}
\label{fig:all}
\end{figure}

  In Figure~\ref{fig:all} we examine a number of variations on our basic model.
Increasing the seed BH mass to $100 \msun$ ({\it upper left panel}) causes a
small increase of the redshift at which a mass of $\sim 10^9 \msun$ can
be achieved. The models in the lower left panel show what happens when
growth by gas accretion is allowed only in halos with $\sigma_{min} >
15 ~\kms$, since it may seem implausible that BHs accrete at the
Eddington rate in halos in which the gas can be easily blown out after
being photoionized. In this case, the Fitchett and Kidder models for
the GW kick velocities can reach a maximum mass of only $\sim 2\times 10^9
\msun$ by $z=6.43$. In the upper right panel, we impose an additional
requirement: halo mergers will lead to the mergers of their central BHs
only if the mass ratio of the two halos, $q$, is between $0.1$ and $1$.
Otherwise the merging halo is assumed to remain an orbiting satellite
(with too long a dynamical friction timescale for spiraling in), so its
BH never reaches the center. In this case the Fitchett and Kidder
models reach a mass of only $3\times 10^8 \msun$ at $z=6.43$, and with
the upper limit to GW kick velocities of Favata \etal the required mass of
$4\times 10^9 \msun$ is barely reached, but for the Favata \etal lower
limit to the GW kick velocities the required mass can still be reached
at $z\simeq 9$.

  We also show the result of raising $\sigma_{min}$ to $50~\kms$ in
the lower right panel, and maintaining the requirement $q>0.1$ on the
halo mass ratio. This higher limit might be more realistic if central
BHs are only fed in galaxies with gravitationally unstable
disks and a sufficient star formation rate (so that the gas does not all
settle in a large, quiescent disk), and where the gravitational
potential is sufficiently deep. Even in this case, the Favata \etal
lower limit to the kick velocity can produce a BH mass above
$4\times 10^9 \msun$ by $z=6.43$, but with higher kick velocities
the BH mass that can be reached is too low.
Note that in the models of \citet{haiman} BHs do not appear at all when
$\sigma < \sigma_{min}$, whereas in our model BHs can still form and
merge (but not accrete gas) for $\sigma<\sigma_{min}$. This, together
with the higher $\vv_{esc}$ we adopt, explains the difference between
Haiman's results and ours.

\section{Conclusions}
\label{sec:conclusions}

  Combining a continued mass growth of BHs from Eddington-limited gas
accretion at high radiative efficiency ($\epsilon \simeq 0.1$) with BH
mergers, we can still account for the presence of BHs with $\sim 10^9
\msun$ at $z=6.43$, starting from stellar black holes of $10 \msun$.
Typically, these BHs can form by the merger of $\sim 10^4$ stellar BH
seeds, growing by a factor $\sim 10^4$ in mass by gas accretion. We have
shown that the ejection of some BHs due to GW recoil does not
necessarily impede the growth of BHs in halo centers.

  Our model for the formation of these BHs involves two optimistic
assumptions: First, that the BH masses grow by accretion with an
e-folding time shorter than $t_{\up Sal}\sim 4\times 10^7$ years, for
most of the time from their birth as stellar black holes at $z\sim 20$
to $z\simeq 6$. Second, that BH mergers in which the recoil velocity is
less than the halo escape velocity result in the BH immediately
returning to the center and continuing its gas accretion, without
significant losses from the energy of the emitted GWs. These two
assumptions are highly questionable. The first one may seem unlikely in
view of the small fraction of galaxies that host quasars at lower
redshift, although this fraction might be high in the earliest massive
halos to form. Regarding the second, a more realistic calculation might
show that black holes spend a lot of time finding their way back to the
halo center after the GW kick, even when they do not escape the halo
\citep{hut,merritt,madau}.

  The earliest redshift at which a mass of $\sim 10^9 \msun$ can be
reached strongly depends on the value of $\epsilon$ assumed. Making
$\epsilon$ slightly greater than $0.1$ would increase $t_{\up Sal}$
and reduce the number of e-folding times available between the epoch
of formation of the first stellar black holes at $z\sim 20$ and the
earliest time when luminous quasars are observed. This would then make
it impossible to attain the required mass. At the same time, that is
perfectly possible if black holes gain their mass in a short period of 
super-Eddington accretion, which is in fact the main alternative to the model
of continuous Eddington-rate growth we have assumed here.

  In the absence of super-Eddington accretion, the scenario of
continuous, luminous gas accretion and large numbers of mergers
implies that quasars should be strongly correlated at high redshift.
The halo hosting the luminous quasar at $z=6.43$ should have
merged from lower mass halos in the recent past. For example, in the
randomly generated merging history used for the realization shown in
Figure 1, the halo mass at $z=9$ is only about one third of its mass
at $z=6.43$, and several halos of mass greater than $10^{11} \msun$
merge with the main halo in the intervening time. Most of these halos
would need to host their own quasar shining near the Eddington
luminosity for most of the time, within a distance of the turn-around
radius of the halo at $z=6.43$ (a few comoving megaparsecs).

\acknowledgments
We are grateful to Zheng Zheng and Adam Steed for discussions. We also
thank Zheng Zheng for the use of his merger tree code. This work was
supported in part by NSF grant NSF-0098515 and NASA grant HST-GO-09838.

\end{document}